\documentclass[prl,aps,graphicx,color]{revtex4}
\usepackage{amsmath}
\usepackage{multicol}
\usepackage{epsfig}

\newcommand{\opr}[1]{\operatorname{#1}}
\newcommand{\Card}[1]{|#1|}
\newcommand{\bea}{\begin{eqnarray}}
\newcommand{\eea}{\end{eqnarray}}
\newcommand{\be}{\begin{equation}}
\newcommand{\ee}{\end{equation}}
\newtheorem{prop}{Proposition}[section]
\newtheorem{definition}[prop]{Definition}

\def\gp{g^\prime}
\def\gpp{g^{\prime\prime}}
\def\lsim{\mathrel{\lower2.5pt\vbox{\lineskip=0pt\baselineskip=0pt
          \hbox{$<$}\hbox{$\sim$}}}}

\title{Privacy Amplification in Quantum Key Distribution:\\
Pointwise Bound {\it versus} Average Bound}
\author{G. Gilbert,$^{\dag}$ M. Hamrick$^{\ddag}$ and F.J. Thayer$^{\ast}$\\
{\em The MITRE Corporation, McLean, Virginia 22102, USA}}

\begin{document}      

\begin{titlepage}
\begin{trivlist}
\item
\vspace*{4.0ex}
{\Large \textsf{MTR 01W0000056}}\\[-0.8ex]
\hrule ~\\[1.8ex]
{\Large \textsf{ MITRE TECHNICAL REPORT}}\\[2.5cm]
\begin{center}
{\huge \textsf{\textbf{Privacy Amplification in Quantum Key Distribution:}}}\\[0.5ex]
{\huge \textsf{\textbf{Pointwise Bound {\it versus} Average Bound}}}\\[3.5cm]
\end{center}
{\Large \textsf{G. Gilbert}}\\[0.8ex]
{\Large \textsf{M. Hamrick}}\\[0.8ex]
{\Large \textsf{F.J. Thayer}}\\[-.4ex]
~\\
{\Large \textsf{\textbf{July 2001}}}\\[3.4cm]
\begin{tabular}{@{\hspace{-0.15in}} l l l l}

{\normalsize \textsf{\textbf{~~~Sponsor:}}} &
{\normalsize \textsf{MITRE \& DOD}} \hspace{2.22in} &
{\normalsize \textsf{\textbf{Contract No.:}}} \phantom{sp} &
{\normalsize \textsf{DAAB07-01-C-C201}} \\[0.2ex]

{\normalsize \textsf{\textbf{~~~Dept. No.:}}} \phantom{sp} &
{\normalsize \textsf{W072}} &
{\normalsize \textsf{\textbf{Project No.:}}}  &
{\normalsize \textsf{51MSR837 \& 0701N020-QC}} \\[0.3cm]

\end{tabular}

\begin{tabular} {@{\hspace{-0.15in}} l l}

\textsf{The views, opinions and/or findings contained in this} &
   \textsf{Approved for public release;} \\

\textsf{report are those of The MITRE Corporation and should not be} \phantom{spospo} &
   \textsf{distribution unlimited.}  \\

\textsf{construed as an official Government position, policy, or} \\

\textsf{decision, unless designated by other documentation.} \\[0.3cm]

\textsf{\copyright 2001 The MITRE Corporation}

\end{tabular}
~\\[0.5cm]

\hspace{-0.2in} {\huge \textsf{\textbf{MITRE}}}\\[0.5ex]
{\large \textsf{\textbf{Washington ${\mathbf C^3}$ Center}}}\\[0.5ex]
{\large \textsf{\textbf{McLean, Virginia}}}\\
\clearpage
\end{trivlist}
\end{titlepage}

\begin{abstract}$\\$~$\\$
In order to be practically useful, quantum cryptography must not only provide a
guarantee of secrecy, but it must provide this guarantee with a useful,
sufficiently large throughput value.
The standard result of generalized privacy amplification yields
an upper bound only on the {\it average value} of the mutual
information available to an eavesdropper. Unfortunately this result
by itself is inadequate for cryptographic applications.
A naive application of the standard result leads one to {\it incorrectly}
conclude that an acceptable upper bound on the mutual information has been achieved.
It is the {\it pointwise value} of the bound on the mutual information, associated with
the use of some specific hash function, that corresponds to actual implementations.
We provide a fully rigorous mathematical derivation that shows how to obtain a
cryptographically acceptable upper bound on the actual, pointwise
value of the mutual information. Unlike the bound on the average
mutual information, the value of the upper bound on the pointwise mutual
information and the number of bits by which the secret key is compressed are specified
by two different parameters,
and the actual realization of the bound in the pointwise
case is necessarily associated with a specific failure probability.
The constraints amongst these parameters, and the effect of their values on the
system throughput, have not been previously analyzed.
We show that the necessary shortening of the key dictated
by the cryptographically correct, pointwise bound, can still produce
viable throughput rates that will be useful in practice.
\end{abstract}

\maketitle 
\begin{multicols}{2}\raggedcolumns

\section{Introduction}
Quantum cryptography has been heralded as providing an important advance in secret
communications because it provides a guarantee
that the amount of mutual information available to an eavesdropper
can unconditionally be made arbitrarily small.
Any {\it practical} realization of quantum key
distribution that consists only of sifting, error correction and authentication
will allow some information leakage, thus necessitating privacy amplification.
Of course, one might contemplate carrying out privacy
amplification after executing a classical key distribution protocol.
In the absence
of any assumed {\it conditions} on the capability of an eavesdropper, it is not
possible to deduce a provable upper bound on
the leaked information in the classical case, so that the subsequent implementation
of privacy amplification would produce nothing, {\it i.e.,} the ``input"
to the privacy amplification algorithm cannot be bounded, and as a result neither
can the ``output." In the case of quantum key distribution,
however, the leaked information
associated with that string which is the input to the privacy amplification algorithm
can be bounded,
and this can be done in the absence of any assumptions about the capability of an
eavesdropper. This bound is not good enough for cryptography, however. Nevertheless,
this bound on the input allows one to prove a bound on the output of privacy
amplification, so that one deduces a final, unconditional upper bound on the
mutual information available to an eavesdropper. Moreover this bound can be made
arbitrarily small, and hence good enough for cryptography, at the cost of
suitably shortening the final string.
\vskip .075in
\noindent{Except that as usually presented this is not exactly true.}
\vskip .075in
\noindent The above understanding is usually presented in connection with the standard
result of generalized privacy amplification given in \cite{BBCM}, which
applies only to the
{\it average} value of the mutual information. The average is taken with respect
to a set of elements, namely, the $universal_2$ class of hash functions
introduced by Carter and Wegman \cite{WC}. The actual implementation of privacy
amplification, however, will be executed by software and hardware that selects a
{\it particular} hash function. The bound on the average value of the mutual information
does not apply to this situation: it does not directly measure the amount of mutual
information available to an eavesdropper in practical quantum cryptography.

In this paper we calculate cryptographically acceptable pointwise
bounds on the mutual information which can be achieved while still
maintaining sufficiently high throughput rates. In contrast to a
direct application of the privacy amplification result
of \cite{BBCM}, we must
also consider and bound a probability of choosing an unsuitable
hash function and relate this to cryptographic properties of the
protocol and the throughput rate. The relation between average bounds
and pointwise bounds of random variables is not new and follows from
elementary probability theory, as was also noticed in \cite{lutkenhaus-practical}.

\section{Privacy Amplification}

In ideal circumstances, the outcome of a $k$-bit key-exchange protocol
is a $k$-bit key shared between Alice and Bob which is kept secret
from Eve. Perfect secrecy means that from Eve's perspective the shared
key is chosen uniformly from the space of $k$-bit keys. In practice,
one can only expect Eve's probability distribution for the shared key
be close to uniform in the sense that its Shannon entropy is close to
its largest possible value $k$.  Moreover, because quantum
key-exchange protocols implemented in practice {\it inevitably}
leak information to Eve, Eve's distribution of
the key is too far from uniform to be usable for cryptographic
purposes. Privacy amplification is the process of obtaining a nearly
uniformly distributed key in a keyspace of smaller bitsize.

We review the standard assumptions of the underlying probability model
of~\cite{BBCM}:
$\Omega$ is the underlying sample space with probability measure
$\mathbf{P}$.  Expectation of a real random variable $X$ with respect
to $\mathbf{P}$ is denoted $\mathbf{E} X$.  $W$ is a random variable
with key material known jointly to Alice and Bob and $V$ is a random
variable with Eve's information about $W$.  $W$ takes values in some
finite keyspace $\mathcal{W}$. The distribution of $W$ is the function
$\mathbf{P}_{\mathcal{W}}(w) = \mathbf{P}(W = w)$ for $w \in
\mathcal{W}$. Eve's distribution having observed a value $v$ of $V$ is
the conditional probability $\mathbf{P}_{\mathcal{W}}|_{V = v}(w) =
\mathbf{P}(W = w | V = v)$ on $\mathcal{W}$.  In the the discussion
that follows, $v$ is fixed and accordingly we denote Eve's
distribution of Alice and Bob's shared key given $v$ by
$\mathbf{P}_{\mathrm{Eve}}$. $\opr{H}$ and $\opr{R}$ denote Shannon and Renyi
entropies of random variables defined on $\mathcal{W}$ relative to
$\mathbf{P}_{\mathrm{Eve}}$.

\begin{definition} Suppose $\mathcal{Y}$ is a keyspace.  If $\alpha$
is a positive real number, a mapping $\gamma: \mathcal{W} \rightarrow
\mathcal{Y}$ is an $\alpha$ strong uniformizer for Eve's distribution iff
$\opr{H}(\gamma) = \sum_{y \in \mathcal{Y}} \mathbf{P}_{\mathrm{Eve}}(\gamma^{-1}(y))
\log_2 \mathbf{P}_{\mathrm{Eve}}(\gamma^{-1}(y)) \geq \log_2 \Card{\mathcal{Y}} -
\alpha$.
\end{definition}

If $\gamma$ is an $\alpha$ strong uniformizer, then we obtain a bound on the mutual 
information between Eve's data $V$ and the image of the hash transformation $Y$ as 
follows:

\begin{equation}
\label{E:alphastrong}
I(Y,V) = I(Y) - H(Y|V) = \log_2\Card{\mathcal{Y}} - \opr{H}(\gamma) \leq \alpha~.
\end{equation}

\begin{definition}
Let $\Gamma$ be a random variable with values in $\mathcal{Y}^\mathcal{W}$
(space of functions $\mathcal{W} \rightarrow \mathcal{Y}$) which is
conditionally independent of $W$ given $V = v$ i.e.
$
\mathbf{P}(\Gamma = \gamma \mbox{ and }
W = w | {V=v}) = \mathbf{P}(\Gamma = \gamma | {V=v}) \, \mathbf{P}( W = w |
{V=v}).
$
$\Gamma$ is an $\alpha > 0$ average uniformizer for Eve's distribution
iff
\begin{equation}
\mathbf{E}( \opr{H} \Gamma) \geq \log_2
\Card{\mathcal{Y}} - \alpha \, 
\end{equation}
where $ \opr{H} \Gamma = \opr{H} \Gamma(z) = \opr{H}(\Gamma(z))$.
\end{definition}

If $\Gamma$ is an $\alpha$ average uniformizer, the bound is on the mutual 
information averaged over the set $\Gamma$:

\begin{equation}
\label{E:alphaaverage}
I(Y,\Gamma V) = I(Y) - H(Y|\Gamma V) = \log_2\Card{\mathcal{Y}} - \mathbf{E}( \opr{H}
\Gamma) \leq \alpha~.
\end{equation}

Uniformizers are produced stochastically. Notice that by the
conditional stochastic independence assumption, $z$ can be assumed to
vary independently of $w \in \mathcal{W}$ with the law
$\mathbf{P}_{\mathrm{Eve}}$.

\begin{prop}
\label{P:strongresult}
Suppose $\Gamma$ is an $\alpha$ average uniformizer.  Then for every $\beta
> 0$, $\Gamma(\omega)$ is a $\beta$ strong uniformizer for $\omega$ outside a set
of probability $\frac{\alpha}{\beta}$.
\end{prop}
{\sc Proof.} Note that for any $\gamma:\mathcal{W} \rightarrow
\mathcal{Y}$, $\opr{H}\gamma$ is at most $\log_2 \Card{\mathcal{Y}}$.  Thus
$\log_2 \Card{\mathcal{Y}} - \opr{H}\Gamma$ is a nonnegative random
variable. Applying Chebychev's inequality to $\log_2
\Card{\mathcal{Y}} - \opr{H}\Gamma$, it follows that for every $\beta>0$,
\begin{eqnarray*}
\mathbf{P}\bigl( \log_2 \Card{\mathcal{Y}} - \beta \geq
\opr{H}\Gamma\bigr) & \leq &
 \frac{1}{\beta} \mathbf{E}(\log_2 \Card{\mathcal{Y}} -
\opr{H}\Gamma) \\
&  = & \frac{1}{\beta} \bigl( \log_2 \Card{\mathcal{Y}} - \mathbf{E}
(\opr{H}\Gamma) \bigr)  \\ & \leq & \frac{1}{\beta} \alpha.
\end{eqnarray*}
The random variable $\Gamma$ is strongly $\mathrm{universal}_2$ iff for all
$x \neq x' \in X$,
\begin{equation} \mathbf{P}\{z: \Gamma(z)(x) = \Gamma(z)(x')\} \leq
\frac{1}{\Card{\mathcal{Y}}}.
\end{equation}
The following is the main result of~\cite{BBCM}:
\begin{prop} {\bf (BBCM Privacy Amplification)}.
\label{P:averageresult}
Suppose $\Gamma$ is a $\mathrm{universal}_2$ family of mappings
$\mathcal{W} \rightarrow \mathcal{Y}$ conditionally independent of
$W$. Then $\Gamma$ is a
$\frac{2^{\log_2 \Card{\mathcal{Y}} - \opr{R}(X)}}{\ln 2}$ 
average uniformizer for $X$.
\end{prop}

\section{Practical Results}

We will refer to the inequality that provides the upper bound on the average
value of the mutual information as the {\it average privacy amplification bound}, or
APA, and we will refer
to the inequality that provides the upper bound on the actual, or pointwise
mutual information as the {\it pointwise privacy amplifcation bound}, or PPA.

In carrying out privacy amplification we must shorten the key by the number of
bits of information that have potentially been leaked to the
eavesdropper \cite{GH_large}. Having
taken that into account, we denote by $g$ the additional number of bits by which the
key length will be further shortened to assure sufficient secrecy, {\it i.e.}, the
additional bit subtraction amount,
and we refer to $g$ as the {\it privacy amplification subtraction parameter}.
With this definition of $g$, Bennett {\it et al.\/} \cite{BBCM} show as a corollary of 
\ref{P:averageresult} that the set of Carter-Wegman hash functions is an
$2^{-g}/\ln 2$ average uniformizer. We thus have
for the APA bound on $\langle I\rangle$, the average value of the
mutual information, the inequality

\begin{equation}
\label{APA}
\langle I\rangle\equiv I(Y,\Gamma V)\leq {2^{-g}\over\ln 2}~.
\end{equation}

\noindent In the case of APA the quantity $g$ plays a dual role:
in addition to representing the number of additional subtraction bits,
for the APA case $g$ also directly determines the upper bound on the
average of the mutual information. 

In the case of PPA we again employ the symbol $g$ to denote
the number of subtraction bits, as above
for APA, but the upper bound on the pointwise
mutual information is now given in terms of a different quantity $\gp$, which
we refer to as the
{\it pointwise bound parameter}. Also in the case of PPA we need the parameter
$\gpp$, which we refer to as the {\it pointwise probability parameter}, in terms of
which we may define the failure probability $P_f$.
This definition is motivated by \ref{P:strongresult}, 
from which we find that the Carter-Wegman hash functions are $2^{-\gp}/\ln 2$
strong uniformizers except on a set of probability 

\begin{equation}
P_f\equiv {2^{-g}\over\ln 2}{\Big /}{2^{-\gp}\over\ln 2}~.
\end{equation}

\noindent We therefore define the pointwise probability parameter as

\begin{equation}
\label{CE}
\gpp \equiv g - \gp ~.
\end{equation}

\noindent Thus the quantities $g$, $\gp$ and $\gpp$
are not all independent, and are constrained by equation \ref{CE}.
In terms of these parameters we have for the PPA bound on $I$, the actual
value of the mutual information, the inequality

\begin{equation}
\label{PPA}
I\equiv I(Y,V)\leq {2^{-\gp}\over\ln 2}={2^{-\left(g-\gpp\right)}\over\ln 2}
\end{equation}

\noindent where the associated failure probability $P_f$ is given by

\begin{equation}
\label{FP}
P_f=2^{-\gpp}~.
\end{equation}

\noindent The failure probability is
not even a defined quantity in the APA case, but it plays a crucial role in the PPA case.
Thus, the bound on the pointwise mutual information is directly determined by the value
of the parameter $\gp$, with respect to which one finds a tradeoff between $g$,
the number of additional compression bits by which the key is shortened,
and $\gpp$, the negative logarithm of the
corresponding failure probability.

\section{Application of Pointwise Bound}

Operationally, it will usually be the case in practice that end-users of
quantum key distribution systems will be first and foremost
constrained to ensure that a given upper bound on the pointwise
mutual information available to the enemy is realized.

To appreciate the significance of the distinction between the PPA and APA results,
we will consider an illustrative example that shows how reliance on the APA bound
can lead to complete compromise of cryptographic security.
We begin with the APA case.
As noted above, in the case of APA the privacy
amplification subtraction parameter, which we will now denote by $g_{APA}$ to emphasize
the nature of he bound, directly specifies both the upper bound on
$\langle I\rangle$ and also the number of bits by which the key needs to be shortened
to achieve this bound.
Without loss of generality we take the value of the
privacy amplification subtraction parameter to be given by $g_{APA}=30$, which means that,
in addition to the compression by the number of bits of information that were
estimated to have been leaked, the final length of the key will be further shortened by
an additional 30 bits. This results in
an upper bound on the average mutual information given by
$\langle I\rangle\leq 2^{-30}/\ln 2\simeq 1.34\times 10^{-9}$, which
we take as the performance requirement for
this example. While this might appear to be an acceptable
bound, the fact that it applies only to the average of the mutual information of
course means that it is not the quantity we require. 

We turn to the PPA case, with respect to which
we will now refer to the privacy amplification subtraction
parameter as $g_{PPA}$.
In order to discuss the PPA bound we must select
appropriate values amongst $g_{PPA}$, $\gp$ and $\gpp$.
In the APA case discussed above,
the bound on the (average) mutual information and the number of subtraction bits are
both specified by the same parameter $g_{APA}$.
In the PPA case, the number of subtraction bits
and the parameter that specifies the bound on the (pointwise) mutual information are
not the same. To achieve the same value for the upper bound on $I$ as we discussed
for the upper bound on $\langle I\rangle$ above, we must select $\gp=30$ as the value
of the pointwise bound parameter. From eq.(\ref{PPA}) this indeed yields the
required inequality $I\leq 2^{-30}/\ln2\simeq 1.34\times 10^{-9}$.
However, with respect to this requirement on the value on the mutual
information, {\it i.e.}, the required final amount of cryptographic secrecy, there are a
denumerable set (since bits are discrete) of different amounts of compression of
the key that are possible to select, each associated with a corresponding failure
probability, $P_f$, in the form of ordered pairs $\left(g_{PPA},\gpp\right)$ that satisfy
the constraint given by $g_{PPA}=\gp+\gpp$ ({\it cf} eq.(\ref{CE})).

Our starting point was the secrecy performance requirement that must be satisfied.
On the basis of the APA analysis above, one might conclude that in order to achieve
the required secrecy performance constraint it is sufficient to shorten the key by
30 bits. However in the PPA case, satisfying the same performance requirement
{\it and} shortening the key by 30 bits means choosing identical values for the
privacy amplification subtraction parameter ($g_{PPA}=30$) and the pointwise bound
parameter ($\gp=30$). However, we note from eq.(\ref{CE}) that in the case of the PPA
bound, $g_{PPA}$ and $\gp$ become the same only when $\gpp=0$, which corresponds to 100\%
failure probability on the upper bound. This is clearly cryptographically useless!

This example emphasizes the importance of assuring a sufficiently small failure
probability in addition to a sufficiently small upper bound on the mutual information.
As we see from the above example, the APA result provides no information about the
correct number of subtraction bits that are required in order to achieve a specified
upper bound on the pointwise mutual information with a suitable failure probability,
for which it is essential to use the
PPA result instead. In Figure 1 we have plotted the failure probability as a function of
the upper bound on the mutual information, for a family of choices of $g_{PPA}$ values.
Returning to the example discussed above for the APA bound, we see that if we need
to achieve an upper bound on $I$ of about $10^{-9}$, we may do so
with a failure probability
of about (coincidentally) $10^{-9}$, at the cost of shortening the final key by 60 bits:
the secrecy is dictated by the pointwise bound parameter value of $\gp=30$,
which is effected by choosing $g_{PPA}=60$, corresponding to $P_f\simeq 10^{-9}$.
Smaller upper bounds can obviously be obtained, with suitable values of the failure
probability, at the cost of further shortening of the key.

\begin{center}

\epsfig{file=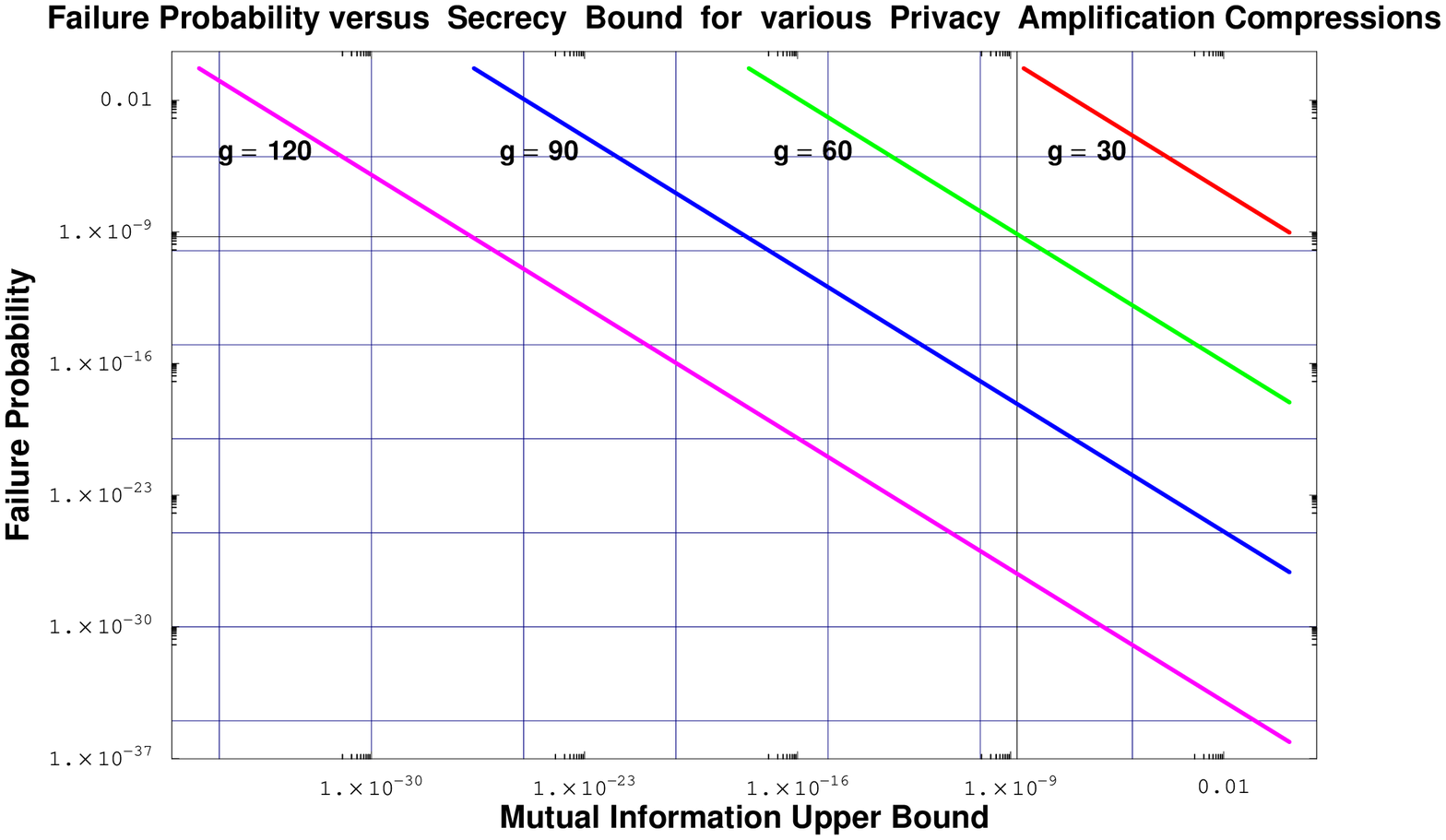,width=9cm}
{\bf Figure 1}

\end{center}

In Figure 2 we plot the throughput of secret Vernam
cipher material in bits per second, as a function of bit cell period, for the
two bit subtraction amounts $g_{PPA}=30$ and $g_{PPA}=60$.
The example chosen is a representative scenario for applied quantum cryptography.
In calculating the rate we follow the method described in reference \cite{GH_large}.
We assume the use of an attenuated, pulsed laser, with Alice located on a low earth orbit
satellite at an altitude of 300 kilometers and Bob located at mean sea level,
with the various system parameters corresponding to those for Scenario ({\it i}) in
Section 5.3.2 in \cite{GH_large}, except that here the source
of the quantum bits operates at a pulse repetition frequency (PRF) of 1 MHz,
and we specifically assume that the enemy does
not have the capability to make use of prior shared entanglement in
conducting eavesdropping attacks.
We see that the additional cost incurred in subtracting
the amount required to achieve the required mutual information bound and failure
probability reduces the throughput rate by an amount that is likely to be acceptable for
most purposes. For instance, for a source PRF of 1 MHz we find that
the throughput rate with a value of $g_{PPA}=30$ is 5614 bits per second. With
a subtraction amount of $g_{PPA}=60$ the throughput rate drops to 5563 bits per second
\cite{blocksize}.

\begin{center}

\epsfig{file=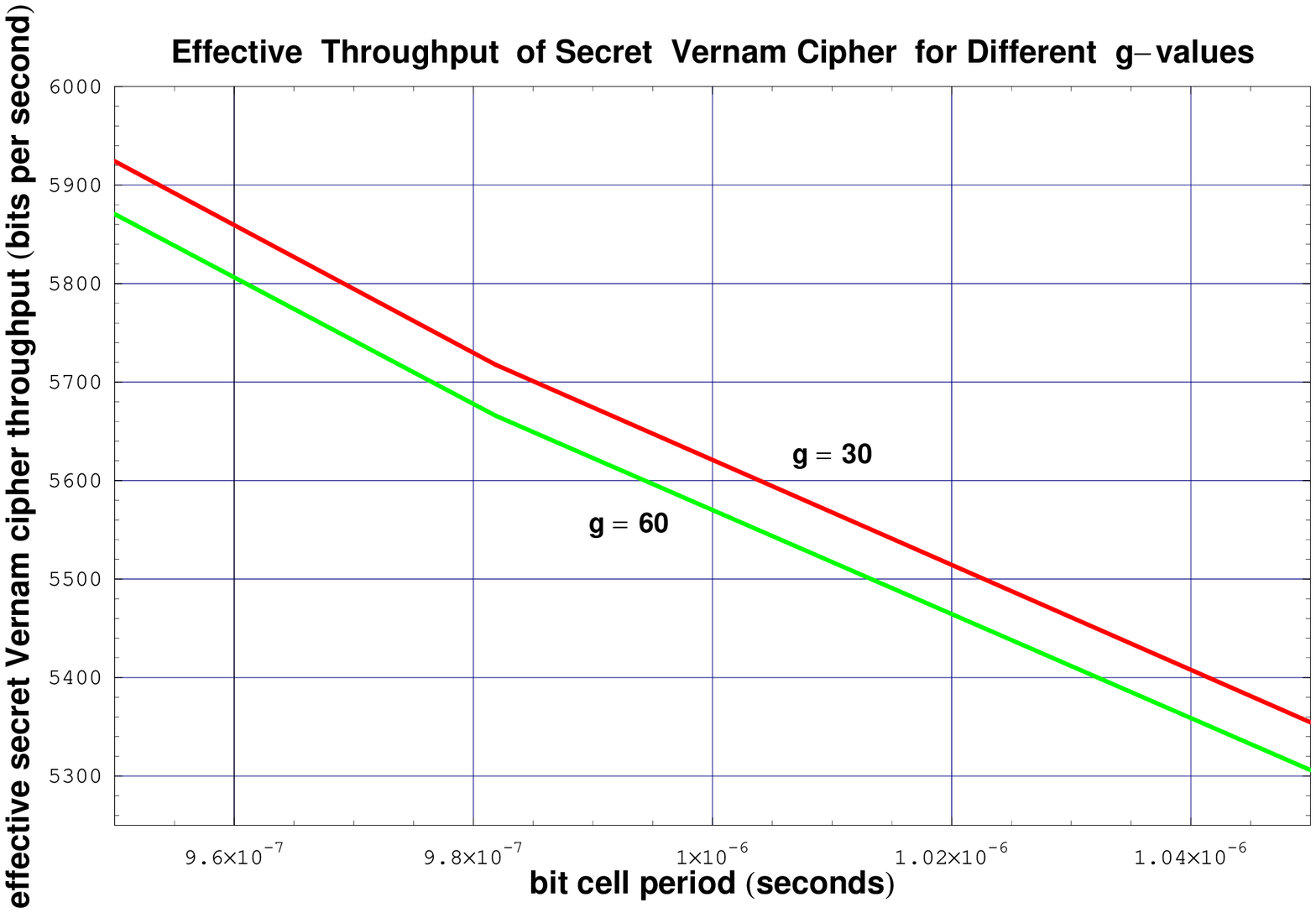,width=9cm}
{\bf Figure 2}

\end{center}

\section{Conclusions}

The significance and proper implementation of privacy amplification in quantum
cryptography are clarified by our analysis.
By itself the bound on the average value of the mutual information presented in
\cite{BBCM} does not allow one to determine the values of parameters required
to bound the actual, pointwise value of the mutual information. Those parameters
must satisfy a constraint, which in turn implies a constraint on
the final throughput of secret key material. We have
rigorously derived the cryptographically meaningful upper bound on the pointwise
mutual information associated with the use of some specific privacy
amplification hash function, and shown that the corresponding requirements on
the shortening of the key still allow viable throughput values.

\vspace*{.175in}
{\footnotesize
$\!\!\!\!\!\!\!\!\!\!$
\dag ~ggilbert@mitre.org\\
\ddag ~mhamrick@mitre.org\\
$\ast$ jt@mitre.org}

\setcounter{unbalance}{15}
\end{multicols}


\begin{thebibliography}{1}
\bibitem{blocksize}
The difference between the two throughput values is about 50 bits per second,
because an additional 30 bits are subtracted per processing block, and in
the example presented there are about 1.6 blocks per second.
See reference \cite{GH_large} for a discussion of processing block size.
\end{thebibliography}
\end{document}